\documentstyle[12pt]{article}

\newcommand{\ra}{\rightarrow}

\newcommand{\s}{\\ \vspace*{-3mm} }
\newcommand{\nn}{\noindent}
\newcommand{\non}{\nonumber}
\newcommand{\beq}{\begin{eqnarray}}
\newcommand{\eeq}{\end{eqnarray}}
\newcommand{\SM}{\mbox{${\cal SM}\ $}}
\newcommand{\SUSY}{\mbox{${\cal SUSY}~$\ }}
\newcommand{\MSSM}{\mbox{${\cal MSSM}$}}

\newcommand{\MeV}{\mbox{Me$\!$V}}
\newcommand{\GeV}{\mbox{Ge$\!$V}}
\newcommand{\TeV}{\mbox{Te$\!$V}}

\begin{document}

\begin{titlepage}

\begin{flushright}
DESY 95--210\\
KA--TP--8--95 \\
hep-ph/9511344 \\
November 1995 \\
\end{flushright}

\vspace{1cm}

\begin{center}
\def\thefootnote{\fnsymbol{footnote}}

{\large\sc {\bf  QCD Corrections to Hadronic Higgs Decays}}

\vspace*{3mm}

\vspace{1cm}

{\sc A.~Djouadi$^{1,2}$\footnote{Supported by Deutsche
Forschungsgemeinschaft DFG (Bonn).}, M. Spira$^{3}$ and P.M.~Zerwas$^2$ }

\vspace{1cm}

$^1$ Institut f\"ur Theoretische Physik, Universit\"at Karlsruhe, \\
D--76128 Karlsruhe, FRG. \\
\vspace{0.3cm}

$^2$ Deutsches Elektronen--Synchrotron DESY, D-22603 Hamburg, FRG. \\
\vspace{0.3cm}

$^3$ II.~Institut f\"ur Theoretische Physik\footnote{Supported by
Bundesministerium f\"ur Bildung und Forschung (BMBF), Bonn, under Contract
05 6 HH 93P (5), and by EU Program {\it Human Capital and Mobility}
through Network {\it Physics at High Energy Colliders} under Contract
CHRX--CT93--0357 (DG12 COMA).}, D-22761 Hamburg, FRG. \\

\vspace{0.3cm}

\end{center}

\vspace{1.6cm}

\begin{abstract}
\normalsize
\noindent

\nn We present an update of the branching ratios for Higgs decays in the
Standard Model and the Minimal Supersymmetric extension of the Standard Model.
In particular, the decays of the Higgs particles to quark and gluon jets are
analyzed and the spread in the theoretical predictions due to uncertainties of
the quark masses and the QCD coupling is discussed.

\end{abstract}

\end{titlepage}

\def\thefootnote{\arabic{footnote}}
\setcounter{footnote}{0}

\setcounter{page}{2}

\subsection*{1. Introduction}

The coupling of the Higgs bosons to other particles grows with the mass of the
particles. This characteristic property is a direct consequence of mass
generation through the Higgs mechanism. To establish the Higgs mechanism
experimentally, it is therefore mandatory to measure the couplings very
accurately \cite{Z1} once scalar particles have been found. The main test
grounds for the Higgs couplings to gauge bosons are the production cross
sections for Higgs-strahlung off gauge bosons and $WW/ZZ$ fusion, and the
widths/branching ratios for Higgs decays to gauge bosons. The Higgs couplings
to heavy quarks determine the cross sections for the production of Higgs
particles in $gg$ fusion at hadron colliders \cite{Z1a,R14A}, as well as the
rate of Higgs bremsstrahlung off heavy quarks at $e^+ e^-$ \cite{Z1b} and
hadron colliders \cite{Z1c}. The measurement of Higgs decay branching ratios,
including $b,c$ quarks and $\tau$ leptons \cite{Z1d}, provides a complementary
method to determine the Higgs couplings. \s

In this note we will reanalyze \cite{GKW} the branching ratios for Higgs
decays to $b,c$ quark jets and to light hadron jets evolving out of
gluon decays,
\begin{eqnarray}
H & \ra & b{\overline{b}}\,/\, c{\overline{c}}\,+\, \dots \label{eq:1} \\
H & \ra & gg\,+\, \dots \label{eq:2}
\end{eqnarray}
The ellipses indicate additional gluon and  quark partons in the final state
due to QCD radiative corrections. Special attention will be paid to
uncertainties related to the $b,c$ quark masses and the QCD coupling
$\alpha_{s}$. It turns out that the evolution of the charm quark mass from low
energy scales, where it can be determined by QCD sum rules, to high energy
scales defined by the Higgs mass, introduces very large uncertainties in the
$c\overline{c}$ branching ratio. The partial width of the second decay mode
(\ref{eq:2})
will be derived for gluon and {\it light} quark final states since heavy quarks
add to the partial width of the first decay process
(\ref{eq:1}). The $b$, $c$ and gluon
decay modes are experimentally important in the Standard Model (${\cal SM}$)
for Higgs masses less than about 150 \GeV. In the minimal supersymmetric
extension (\MSSM) $b$ quark decays may be dominant for a much wider range in
the parameter space.

\subsection*{2. Standard Model}
%        ==============
\subsubsection*{2.1 b, c quark decays of the \SM Higgs particle}
%           =================================================
The particle width for decays to (massless) $b,c$ quarks directly coupled to
the \SM Higgs particle is given, up to ${\cal O}(\alpha_{s}^{2})$ QCD radiative
corrections\footnote{The effect of the electroweak radiative corrections in
the branching ratios is negligible \cite{helw}.}, (Fig.1a) by the well-known
expression
\cite{Z3,Z3A,chetyr}
\begin{equation}
\Gamma [H\, \ra \, Q{\overline{Q}}] = \frac{3 G_F M_H } {4\sqrt{2}\pi}
\ \overline{m}_Q^2(M_H)\ [\Delta_{\rm QCD}+\Delta_t]
\label{eq:hqq}
\end{equation}
\begin{eqnarray*}
\Delta_{\rm QCD} & = & 1 + 5.67 \frac{\alpha_s (M_H)}{\pi} + (35.94 - 1.36 N_F)
\left( \frac{\alpha_s (M_H)}{\pi} \right)^2 \\
\Delta_t & = & \left(\frac{\alpha_s (M_H)}{\pi}\right)^2 \left[ 1.57 -
\frac{2}{3} \log \frac{M_H^2}{M_t^2} + \frac{1}{9} \log^2
\frac{\overline{m}_Q^2 (M_H)}{M_H^2} \right] \non
\end{eqnarray*}
in the ${\overline{\rm MS}}$ renormalization scheme; the running quark
mass and the QCD coupling are defined at the scale of the Higgs mass,
absorbing this way any large logarithms. The quark masses can be neglected
in general except for top quark decays where
this approximation holds only sufficiently far above threshold; the QCD
correction in this case are given in the Appendix.  \s

Since the relation between the pole mass $M_{c}$ of the charm quark and
the ${\overline{\rm MS}}$ mass evaluated at the pole mass
${\overline{m}}_{c}(M_{c})$ is badly convergent \cite{Z5}, we will adopt
the running quark masses ${\overline{m}}_{Q}(M_{Q})$ as starting points.
They have been extracted directly from QCD sum rules evaluated in a
consistent ${\cal O}(\alpha_{s})$ expansion \cite{Z4}. The evolution
from $M_{Q}$ upwards to a renormalization scale $\mu$ is given by
\begin{eqnarray}
{\overline{m}}_{Q}\,(\mu )&=&{\overline{m}}_{Q}\,(M_{Q})
\,\frac{c\,[\alpha_{s}\,(\mu)/\pi ]}{c\, [\alpha_{s}\,(M_{Q})/\pi ]}
\label{eq:msbarev}
\end{eqnarray}
with \cite{Z3A,Z5}
\begin{eqnarray*}
c(x)&=&(\frac{25}{6}\,x)^{\frac{12}{25}} \, [1+1.014x+1.389\,x^{2}]
\hspace{1.0cm} \mbox{for} \hspace{.2cm} M_{c}\,<\mu\,<M_{b}\\
c(x)&=&(\frac{23}{6}\,x)^{\frac{12}{23}} \, [1+1.175x+1.501\,x^{2}]
\hspace{1cm} \mbox{for} \hspace{.2cm} M_{b}\,<\mu
\end{eqnarray*}
For the charm quark mass the evolution is determined by eq.(\ref{eq:msbarev})
up to the scale $\mu=M_b$, while for scales above the bottom mass the
evolution must be restarted at $M_Q = M_b$. Typical values of the
running $b,c$ masses at the scale $\mu = 100$ \GeV, characteristic for
the Higgs mass, are displayed in Table 1. The evolution has been
calculated  for the QCD coupling
\begin{eqnarray}
\alpha_{s}(M_{Z})&=&0.118\,\pm \, 0.006
\end{eqnarray}
defined at the $Z$ mass \cite{Z6A}. The large uncertainty in the running
charm mass is a consequence of  the small scale at which the evolution starts
and where the errors of the QCD coupling are very large.
\begin{table}[hbt]
\begin{center}
\begin{tabular}{|c|c|cc|c|} \hline
& & & & \\
&$ \alpha_{s}(M_{Z}) $
& $ {\overline{m}}_{Q}(M_{Q})$\ & \ $M_{Q}\,=\,M_{Q}^{pt2} $
& \ ${\overline{m}}_{Q}\,(\mu\,=\ 100~\GeV) $ \\
& & & & \\ \hline
& & & & \\
$b$ & $0.112$ & $(4.26 \pm 0.02)~\GeV$ & $(4.62 \pm 0.02)~\GeV$
& $(3.04 \pm 0.02)~\GeV$ \\
    & $0.118$ & $(4.23 \pm 0.02)~\GeV$ & $(4.62 \pm 0.02)~\GeV$
& $(2.92 \pm 0.02)~\GeV$ \\
    & $0.124$ & $(4.19 \pm 0.02)~\GeV$ & $(4.62 \pm 0.02)~\GeV$
& $(2.80 \pm 0.02)~\GeV$ \\
& & & & \\ \hline
& & & & \\
$c$ & $0.112$ & $(1.25 \pm 0.03)~\GeV$ & $(1.42 \pm 0.03)~\GeV$
& $(0.69 \pm 0.02)~\GeV$ \\
    & $0.118$ & $(1.23 \pm 0.03)~\GeV$ & $(1.42 \pm 0.03)~\GeV$
& $(0.62 \pm 0.02)~\GeV$ \\
    & $0.124$ & $(1.19 \pm 0.03)~\GeV$ & $(1.42 \pm 0.03)~\GeV$
& $(0.53 \pm 0.02)~\GeV$ \\
& & & & \\ \hline
\end{tabular}
\end{center}
\caption[]{\it The running $b,c$ quark masses in the ${\overline{\rm
MS}}$ renormalization scheme at the scale $\mu = 100$ \GeV. The starting
points ${\overline{m}}_{Q}(M_{Q})$ of the evolution are extracted from
QCD sum rules \cite{Z4}; the pole masses $M_{Q}^{pt2}$ are defined by
the ${\cal O}(\alpha_{s})$ relation ${\overline{m}}_{Q}(M_{Q}^{pt2})=
M_{Q}^{pt2}/[1+4\alpha_{s}/3\pi]$ with the running masses.}
\end{table}

\subsubsection*{2.2 Higgs decay to light hadron jets}
%           ================================
The decay of the Higgs boson to gluons is mediated by heavy quark loops
in the Standard Model (Fig.1b); the partial decay width \cite{R11A} is given by
\begin{eqnarray}
\Gamma_{LO}\, [H\ra gg] = \frac{G_{F}\, \alpha_{s}^{2}\,M_{H}^{3}}
{36\,\sqrt{2}\,\pi^{3}} \left| \sum_{t,b} A^H(\tau_Q) \right|^2
\label{eq:hgglo}
\end{eqnarray}
with the form factor
\begin{eqnarray*}
A^H (\tau) & = & {\textstyle \frac{3}{2}} \tau \left[ 1+(1-\tau) f(\tau)
\right] \\
f(\tau) & = & \left\{ \begin{array}{ll}
\displaystyle \arcsin^2 \frac{1}{\sqrt{\tau}} & \tau \ge 1 \\
\displaystyle - \frac{1}{4} \left[ \log \frac{1+\sqrt{1-\tau}}
{1-\sqrt{1-\tau}} - i\pi \right]^2 & \tau < 1
\end{array} \right.
\end{eqnarray*}
The parameter $\tau_Q= 4M_Q^2/M_H^2$ is defined by the pole mass of the
heavy loop quark $Q$.
For large quark masses the form factor approaches unity. QCD radiative
corrections are built up by the exchange of virtual gluons and the
splitting of a gluon into two gluons or a quark--antiquark pair,
Fig.1b. If all quarks $u, \cdots,b$ are
treated massless at the renormalization scale $\mu \sim M_{H} \sim 100$ \GeV,
the radiative corrections can be approximated very well by \cite{Hgg}
\begin{eqnarray}
&& \Gamma^{N_F}\,[H\ra gg\,(g),\,q{\overline{q}}g]= \Gamma_{LO}\,
\left[\alpha_{s}^{(N_{F})}(\mu )\right]\,\left\{1+E^{N_{F}}
\frac{\alpha_{s}^{(N_{F})}(\mu)}{\pi}\ \right\} \label{eq:hgg} \\
&& \hspace*{1.3cm}
E^{N_{F}}=\frac{95}{4}-\frac{7}{6}N_{F} \,+\,\frac{33-2\,N_{F}}
{6}\log \,\frac{\mu ^{2}}{M_{H}^{2}}\nonumber
\end{eqnarray}
with $N_{F}=5$ light quark flavors. The radiative corrections are very
large, nearly doubling the partial width. \s

The final states $H\,\ra\, b{\overline{b}}g$ and $c{\overline{c}}g$ are also
generated through processes in which the $b,c$ quarks are coupled to the Higgs
boson directly (Fig.~1a). Gluon splitting $g\,\ra\, b{\overline{b}}$ in
$H\,\ra\,gg$ (Fig.~1b) increases the inclusive decay probabilities\footnote{The
two contributions add up incoherently in the limit where the final state quark
masses are neglected [apart from the Yukawa Higgs couplings]. The topology of
the final states is in general different for decays to $b,c$ quarks through the
direct couplings or gluon splitting: in the former decay mode, quark and
anti--quark jets are emitted primarily back--to--back, while in the latter
decay mode they tend to be collinear.} $\Gamma(H \ra b\bar{b}+ \dots)$ {\it
etc.} Since $b$ quarks, and eventually $c$ quarks, can in principle be tagged
experimentally, it is physically meaningful to consider the particle width of
Higgs decays to gluon and light $u,d,s$ quark final jets separately. The
contribution of $b,c$ quark final states to the coefficient $E^{N_{F}}$ in
eq.(\ref{eq:hgg}) is given by
\begin{eqnarray}
\delta \, E^{b,c}&=&\,-\frac{7}{3}\,+\,
\frac{1}{3}\,\left[\,\log \, \frac{M_{H}^{2}}
{M_{b}^{2}}\, + \, \log \,\frac{M_{H}^{2}}
{M_{c}^{2}}\right]\nonumber
\end{eqnarray}
in the limit $M_{H}^{2}\, \gg \, M_{b,c}^{2}$.
Instead of naively subtracting this contribution,
it may be noticed that the mass logarithms can
be absorbed by changing the number of active
flavors from $N_{F}=5$ to $N_{L}=3$ in the
QCD coupling,
\begin{eqnarray}
\alpha_{s}^{(5)}\,(\mu)&=&\alpha_{s}^{(3)}\,
(\mu)\,\left\{1+\frac{\alpha_{s}^{(3)}}
{6\,\pi}\left[\log \frac{\mu^{2}}{M_{c}^{2}}+
\log \frac{\mu^{2}}{M_{b}^{2}}\right]+
\dots \right\}\nonumber
\end{eqnarray}
This way we arrive again at an equally simple expression
\begin{eqnarray}
&& \Gamma [H\ra gg(g), q{\overline{q}} g ] = \Gamma_{LO} \left[
\alpha_{s}^{(N_{L})} (\mu) \right] \left\{ 1 +  E^{N_{L}} \frac{\alpha_{s}
^{(N_{L})} (\mu)} {\pi} \right\} \\
&& \hspace*{1.3cm} E^{N_{L}}
= \frac{95}{4} - \frac{7}{6} N_{L} + \frac{33-2N_{L}}{6}
\log \frac{\mu^{2}} {M_{H}^{2}} \nonumber
\end{eqnarray}
for $N_{L}=3$ light $q=u,d,s$ quark flavors in the final state. \s

The subtracted parts may be added to the partial decay widths into $c$
and $b$ quarks. Up to ${\cal O}(\alpha_s^3)$, they are given by the
difference of the gluonic widths
[eq.(\ref{eq:hgg})] for the corresponding number of flavors $N_F$,
\begin{eqnarray}
\delta \Gamma [H\to c\bar c+\dots] & = & \Gamma^{4} - \Gamma^{3}
\nonumber \\
\delta \Gamma [H\to b\bar b+\dots] & = & \Gamma^{5} - \Gamma^{4}
\end{eqnarray}
The ${\overline{\rm MS}}\ \Lambda $ parameters for three, four and five
flavors in the QCD coupling
\begin{equation}
\alpha_s^{(N_F)} (\mu) = \frac{12\pi}{(33-2N_F)
\log(\mu^2/\Lambda_{N_F}^2)}~ \left[ 1 - 6~\frac{153-19N_F}{(33-2N_F)^2}~
\frac{\log\log(\mu^2/\Lambda_{N_F}^2)} {\log(\mu^2/\Lambda_{N_F}^2)}
\right]
\end{equation}
are given in Table 2 together with the effective couplings $\alpha_s^{(N_F)}
(M_Z)$ for three, four and five flavors corresponding to the QCD scales
$\Lambda_{N_F}$.  In $\alpha_s^{(4)}(M_Z)$ the contribution of the $b$ quark
is subtracted and in $\alpha_s^{(3)} (M_Z)$ the contributions of both
the $b$ and $c$ quarks. \s

\begin{table}[hbt]
\begin{center}
\begin{tabular}{|c|c||c|c||c|c|} \hline
& & & & & \\
$\alpha_s^{(5)} (M_Z)$ & $\Lambda_{5} [\MeV]$ &
$\alpha_s^{(4)} (M_Z)$ & $\Lambda_{4} [\MeV]$ &
$\alpha_s^{(3)} (M_Z)$ & $\Lambda_{3} [\MeV]$ \\
& & & & & \\ \hline \hline
& & & & & \\
0.112 & 159 & 0.107 & 238 & 0.101 & 286 \\
0.118 & 226 & 0.113 & 327 & 0.105 & 378 \\
0.124 & 312 & 0.118 & 434 & 0.110 & 483 \\
& & & & & \\ \hline
\end{tabular}
\end{center}
\caption[]{\it QCD scales $\Lambda_{N_F}$ for the range of uncertainty in
the strong coupling constant $\alpha_s^{(5)} (M_Z) = 0.118 \pm 0.006$. The
effective couplings $\alpha_s^{(4)} (M_Z)$, with the $b$ quark
decoupled, and $\alpha_s^{(3)}(M_Z)$, with both the $b$ and $c$ quarks
decoupled, are given for comparison, too.}
\end{table}

With $E^{3}\,=\,20.25$, the QCD radiative corrections still amount\footnote{If
the $b,c$ quarks are included in the final state, the partial width increases
by an additional 20\%.} to $\sim 70\%$. However, a large fraction of the
corrections can be absorbed by choosing, for the proper renormalization scale,
the BLM scale \cite{BLM} which maps contributions associated with gluon
self-energies into the effective QCD coupling; this is technically implemented
by choosing $\mu$ such that the coefficient of the $N_F$ or $N_{L}$ term
vanishes:
\begin{eqnarray}
\mu_* &=&e^{-\frac{7}{4}}\,M_{H}\,\approx \, 0.17\,M_{H}\nonumber
\end{eqnarray}
The QCD corrections to the partial width
\begin{eqnarray}
\Gamma [H\ra gg(g), q{\overline{q}}g ]&=&\Gamma_{LO}\left[\,
\alpha_{s}^{(N_F)}(\mu_{*})\right]\, \left\{ 1\,+\,\frac{9}{2}
\frac{\alpha_{s}^ {(N_F)}(\mu_{*})}{\pi}\right\}
%\nonumber
\end{eqnarray}
are reduced in this approach to a comfortable level of 15 to 25\%.

\subsubsection*{2.3 Numerical evaluation}
%               ========================
The numerical analysis of the branching ratios for the Higgs decays in
the Standard Model has been performed for the set of parameters given
in the tables and the top quark mass\footnote{This value of the
top quark mass $M_t$
corresponds to the average of all measurements at the Tevatron presented
in Ref.\cite{R8A}.}
\begin{eqnarray*}
%\alpha_{s}^{(5)}\,(M_{Z}^{2})&=&0.118\,\pm \, 0.006 \\
M_{t}&=&(176\, \pm \, 11)~\GeV
%M_{b}^{pt2}&=&(4.62\, \pm \, 0.02)~\GeV \\
%M_{c}^{pt2}&=&(1.42\, \pm \, 0.03)~\GeV
\end{eqnarray*}
To estimate systematic uncertainties, the variation of the $c$ mass has
been stretched over $2\sigma$ and the uncertainty of the $b$ mass to
0.05 \GeV. However, the dominant error in the predictions is due to the
uncertainty in $\alpha_{s}$ which migrates to the running quark masses
at the high energy scales. \s

The results for the branching ratios are displayed in Fig.2. Separately shown
are the branching ratios for $\tau$'s, $c,b$ quarks, gluons plus light quarks
and electroweak gauge bosons. The uncertainties in the prediction for the charm
and gluon branching ratios are very large. Increasing $\alpha_{s}$ reduces the
value of the running $c$ mass quite dramatically\footnote{The value of $BR_c$
is significantly smaller in Ref.\cite{GKW} for two reasons. (i) The ratio
between the $\overline{\rm MS}$ mass $\overline{m}_c (M_c)$ and the
pole mass $M_c^{pt3}$ to ${\cal O}(\alpha_s^2)$ is smaller than the
corresponding ratio for $M_c^{pt2}$ to ${\cal O}(\alpha_s)$.
However, since $ \overline{m}_c(M_c)$ has been determined through QCD sum rules
only to ${\cal O}(\alpha_s)$, the parameter $M_c=M_c^{pt2}$ is the proper pole
mass to be used in a consistent analysis up to ${\cal O}(\alpha_s)$.
We have performed the evolution of the running $\overline{\rm MS}$ mass
with and without the ${\cal O}(\alpha_s^2)$ contribution; the difference
between the two results at the scale of the Higgs mass turned out to be
negligible. The present analysis is therefore theoretically consistent.
(ii) Moreover, in Ref.\cite{GKW} the average LEP $\alpha_{s}$ value has been
adopted which is larger than the world average value including
deep--inelastic scattering data. This gives rise to a faster fall--off of the
running charm mass $ \sim [\alpha_s(\mu)]^{12/23}$ at large scales.}. \s

Nevertheless, the expected hierarchy of the Higgs decay modes is clearly
visible in Fig.2 despite these uncertainties. $BR_{\tau}$ is more than an order
of magnitude smaller than $BR_{b}$, a straight consequence of the ratio between
the two masses squared and the color factor. As a result of the small charm
quark mass at the scale of the Higgs mass, the ratio of $BR_{c}$ to $BR_{b}$ is
reduced by much more than an order of magnitude, which would have been naively
expected. Taking these subtle QCD effects into account, the measurement of the
decay branching ratios provides an excellent method to explore the physical
nature of the Higgs particle.

\subsection*{3. The Minimal Supersymmetric Standard Model}
%            ============================================
We have performed a similar analysis for the hadronic decay modes of the
Higgs bosons $h,\, H,\, A,\,H^{\pm}$ in the Minimal Supersymmetric
extension of the Standard Model (\MSSM). Apart from the usual modifications
$g_{Q}^{i}$ of the couplings, the analytic expressions for the
partial widths of the scalar neutral Higgs bosons $h,H$ are the same as
in the Standard Model, eqs.(\ref{eq:hqq}) and (\ref{eq:hgg}). In the
massless quark limit [the general massive case is treated in the Appendix],
the QCD radiatively corrected decay widths into quarks are given by
\begin{equation}
\Gamma [\Phi \, \ra \, Q{\overline{Q}}] =
\frac{3G_F M_\Phi }{4\sqrt{2}\pi} \overline{m}_Q^2 (g_Q^\Phi)^2
\left[ \Delta_{\rm QCD} + \Delta_t^\Phi \right]
\end{equation} \vspace*{-0.4cm}
\begin{eqnarray}
\Delta_t^{h/H} & =& \frac{g_t^{h/H}}{g_Q^{h/H}}~\left(\frac{\alpha_s
(M_{h/H})}{\pi}
\right)^2 \left[ 1.57 - \frac{2}{3} \log \frac{M_{h/H}^2}{M_t^2}
+ \frac{1}{9} \log^2 \frac{\overline{m}_Q^2 (M_{h/H})}{M_{h/H}^2}\right]\non \\
\Delta_t^A & = & \frac{g_t^A}{g_Q^A}~\left(\frac{\alpha_s (M_A)}{\pi} \right)^2
\left[ 3.83 - \log \frac{M_A^2}{M_t^2} + \frac{1}{6} \log^2
\frac{\overline{m}_Q^2 (M_A)}{M_A^2} \right] \non
\end{eqnarray}
\begin{equation}
\Gamma [\,H^{+} \ra \, U{\overline{D}}\,] =
\frac{3 G_F M_{H^\pm}}{4\sqrt{2}\pi} \, \left| V_{UD} \right|^2 \,
\left[ \overline{m}_U^2 (g_U^{A})^2 + \overline{m}_D^2 (g_D^{A})^2 \right]
\Delta_{\rm QCD}
\label{eq:hcud}
\end{equation}
with $\Phi=h,H,A$. [Eq.(\ref{eq:hcud}) is valid if either the first or the
second term is dominant.] The relative couplings $g_{Q}^{\Phi}$ have recently
been collected in Ref.\cite{susycoup}; the masses in the Yukawa couplings
are to be evaluated at the scales $M_\Phi$ and $M_{H^\pm}$. \s

Since the $b$ quark couplings to the Higgs bosons may be strongly enhanced and
the $t$ quark couplings suppressed in the \MSSM, $b$ loops can contribute
significantly to the $gg$ coupling so that the approximation $M_{Q}^{2} \gg
M_{H}^{2}$ cannot be applied any more in general. Nevertheless, it turns out
{\it a posteriori} that this is an excellent approximation for the QCD
corrections. The $LO$ width for $h,H \ra gg$ is obtained from
eq.(\ref{eq:hgglo}) by substituting $A^H \ra g_Q^{h,H} A^{h,H}$; for the
pseudoscalar Higgs decays, we find \cite{R14A}
\begin{eqnarray}
\Gamma_{LO}\,[A\ra gg]&=&\frac{G_{F}\,
\alpha_{s}^{2}\,M_{A}^{3}}{16\,\sqrt{2}\,
\pi^{3}} \left| \sum_{t,b} g_Q^A A^A(\tau_Q) \right|^2 \\
A^A (\tau) & = & \tau f(\tau) \nonumber \\
\Gamma [\,A\,\ra \, gg(g),q{\overline{q}}g\,] &=&
\Gamma_{LO}\,\left[\alpha_{s}^{(N_{F})}(\mu
)\right]\,\left\{1+E^{N_{F}}\frac{\alpha_{s}^
{(N_{F})}(\mu)}{\pi}\ \right\} \\
E^{N_{F}}&=&\frac{97}{4}-\frac{7}{6}N_{F}
\,+\,\frac{33-2\,N_{F}}{6}\log \,\frac{\mu
^{2}}{M_{A}^{2}}\nonumber
\end{eqnarray}

To illustrate the size of the uncertainties introduced into the predictions
by the QCD parameters, the branching ratios have been calculated
for a specific set of parameters. The top mass is varied within $M_{t}=(176
\pm 11)$ \GeV. In addition to the other parameters defined in the previous
section, the running mass of the $s$ quark at the scale 1 \GeV~and the
CKM type mixing parameter $V_{cb}$ are chosen as
\begin{eqnarray*}
&& {\overline{m}}_{s}\,(1~\GeV)=(0.190\,\pm\,0.040) \GeV \\
&& |V_{cb}| = 0.040 \pm 0.008
\end{eqnarray*}
while the \SUSY parameters are set to
\begin{eqnarray*}
&& \tan \beta = 1.6 \\
&& \mbox{\SUSY scale parameter: }\,M_{S}=\,1~\TeV \\
&& \mbox{``typical" mixing: } A_{t}=\,-\mu \,=\, 1~\TeV
\end{eqnarray*}
\SUSY masses and couplings have been calculated according to the
RG program described
in Ref.\cite{carena}. Varying the \SUSY parameters does not change the picture
of the QCD corrections and the uncertainties exemplified for the set of
parameters chosen above. The final results\footnote{The Fortran code for the
partial decay widths in the \SM and \MSSM~may be obtained from
djouadi@desy.de or spira@desy.de.} are displayed in Figs.3a--d. The
branching ratios are separated again for final states including $b,c$ quarks
[labeled $b\bar b$ and $c\bar c$] and gluons plus light quarks [labeled $gg$].
The main sources of uncertainties
for the branching ratios are the charm and gluon decays. The branching ratios
for $b$ and $\tau$ decays are less affected by $\alpha_{s}$.
The uncertainty in the top quark mass affects primarily
the upper limit of the light CP--even scalar mass $M_h$ and the couplings at
the electroweak level, as shown in Fig.3a. The uncertainty in $M_h$ migrates to
the partial widths of the heavy Higgs bosons through cascade decays; this
second step is indicated by the error bars attached to the curves in
Figs.~3b--d.

\vspace*{1cm}

\nn {\bf Acknowledgements} \s

\nn We are very grateful for discussions with A.\,Kataev, Y.\,Okada and
M.\,Peskin. We thank M.\,Carena for providing us with the Fortran code for
calculating the \SUSY Higgs parameters. A.D. thanks the Theory Group for
the warm hospitality extended to him at DESY.

%\newpage
\vspace*{1cm}

\renewcommand{\theequation}{A.\arabic{equation}}
\setcounter{equation}{0}

\subsection*{APPENDIX}

For completeness, we present in this Appendix the expressions of the
leading order QCD corrections to \SM and \MSSM~Higgs boson decays
involving non--zero mass effects for heavy quarks. As a general example we
will consider top quarks. \s

The partial decay widths of the \underline{CP--even Higgs bosons}
$\Phi=H_{\rm SM},h$ and $H$ into top quark pairs,
in terms of the top quark {\it pole} mass,
is given by
\begin{eqnarray}
\Gamma [\Phi \ra t\bar{t}\,]= \frac{3G_F M_\Phi}{4 \sqrt{2} \pi} \,
(g_t^\Phi)^2 \, m_t^2 \, \beta^3 \, \left[ 1 +\frac{4}{3} \frac{\alpha_s}{\pi}
\Delta^{\Phi} \right]
\end{eqnarray}
where $\beta = (1-4m_t^2/M_\Phi^2)^{1/2}$ is the velocity of the top quarks.
To leading order, the QCD correction factor is given by \cite{Z3,Drehi,R15}
\begin{eqnarray}
\Delta^{\Phi} = \frac{1}{\beta}A(\beta) + \frac{1}{16\beta^3}(3+34\beta^2-
13 \beta^4)\log \frac{1+\beta}{1-\beta} +\frac{3}{8\beta^2}(7 \beta^2-1)
\end{eqnarray}
with
\begin{eqnarray}
A(\beta) &= & (1+\beta^2) \left[ 4 {\rm Li}_2 \left( \frac{1-\beta}{1+\beta}
\right) +2 {\rm Li}_2 \left( -\frac{1-\beta}{1+\beta} \right) -3 \log
\frac{1+\beta}{1-\beta} \log \frac{2}{1+\beta} \right. \non \\
& & \left. -2 \log \frac{1+\beta}{1-\beta} \log \beta \right] -
3 \beta \log \frac{4}{1-\beta^2} -4 \beta \log \beta
\non
\end{eqnarray}
[Li$_2$ is the Spence function defined by Li$_2(x)= -\int_0^x dy y^{-1}
\log(1-y)$.] \s

The partial decay width of the \underline{CP--odd Higgs boson} $A$ into top
quark pairs reads correspondingly
\begin{eqnarray}
\Gamma [A \ra t\bar{t}\,]= \frac{3G_F M_A}{4 \sqrt{2} \pi} \, (g_t^A)^2 \,
m_t^2 \, \beta \, \left[ 1 +\frac{4}{3} \frac{\alpha_s}{\pi} \Delta^{A} \right]
\end{eqnarray}
where the QCD correction factor is given by \cite{Drehi,R15}
\begin{eqnarray}
\Delta^{A} = \frac{1}{\beta}A(\beta) + \frac{1}{16\beta}(19+2\beta^2+
3 \beta^4)\log \frac{1+\beta}{1-\beta} +\frac{3}{8}(7 -\beta^2)
\end{eqnarray}

The partial decay width of the \underline{charged Higgs particles} decay
into top and bottom quarks may be written
\begin{eqnarray}
\Gamma [H^+ \rightarrow t \bar{b}\,] &=&  \frac{3G_F M_{H^\pm}}{4 \sqrt{2} \pi}
\, \lambda^{1/2} \, \left\{ (1-\mu_t -\mu_b) \left[ \frac{m_t^2} { {\rm tg}^2
\beta } \left( 1+ \frac{4}{3} \frac{\alpha_s}{\pi} \Delta_{tb}^+ \right)
\right. \right. \\
&& \left. \left. +m_b^2 {\rm tg}^2 \beta \left( 1+ \frac{4}{3} \frac{\alpha_s}
{\pi} \Delta_{bt}^+ \right) \right]
-4m_tm_b \sqrt{\mu_t \mu_b} \left( 1+ \frac{4}{3}
\frac{\alpha_s}{\pi} \Delta_{tb}^- \right) \right\} \non
\end{eqnarray}
where $\mu_t=m_t^2/M_{H^\pm}^2$, $\mu_b=m_b^2/M_{H^\pm}^2$ and $\lambda$ the
usual two--body phase space function $\lambda= (1-\mu_t-\mu_b)^2-4 \mu_t
\mu_b$;
again the quark masses are the pole masses. In the approximation where the
$b$ quark mass is neglected the QCD factors $\Delta_{ij}^\pm$ are given by
\cite{R15}
\begin{eqnarray}
\Delta_{ij}^{+} &=&  \frac{9}{4} + \frac{ 3-2\mu_i+2\mu_j}{4} \log
\frac{\mu_i}{\mu_j} + \frac{ (\frac{3}{2}-\mu_i-\mu_j) \lambda+5 \mu_i
\mu_j}{2 \lambda^{1/2} (1-\mu_i -\mu_j)} \log x_i x_j  +B_{ij} \non \\
\Delta_{ij}^{-} &=&  3 + \frac{ \mu_j-\mu_i}{2} \log \frac{\mu_i}{\mu_j}
+ \frac{ \lambda +2(1-\mu_i-\mu_j)} { 2 \lambda^{1/2} } \log x_i x_j +B_{ij}
\end{eqnarray}
with $x_i= 2\mu_i/[1-\mu_i-\mu_j+\lambda^{1/2}]$ and
\begin{eqnarray*}
B_{ij} &=& \frac{1-\mu_i-\mu_j} { \lambda^{1/2} } \left[ 4{\rm Li_{2}}(x_i
x_j)- 2{\rm Li_{2}}(x_i) -2{\rm Li_{2}}(x_j) +2 \log x_i x_j \log (1-x_ix_j)
\right. \non \\
&& \left. \hspace*{2.2cm} - \log x_i \log (1-x_i) - \log x_j \log (1-x_j)
\right] \non \\
&& - 4 \left[ \log (1-x_i x_j)+ \frac{x_i x_j}{1-x_i x_j} \log x_i x_j
\right]\non
\\ && +\frac{ \lambda^{1/2}+\mu_i-\mu_j } {\lambda^{1/2} } \left[
\log (1-x_i) +\frac{x_i}{1-x_i} \log x_i \, \right] \non \\
&& +\frac{\lambda^{1/2}-\mu_i+\mu_j} { \lambda^{1/2} } \left[
\log (1-x_j) +\frac{x_j}{1-x_j} \log x_j \right]
\end{eqnarray*}

Well above the $t\bar{t}$ threshold, the QCD factors $\Delta^+_{ij}$ reduce to
\begin{eqnarray}
\Delta^+_{tb} = \frac{9}{4} + \frac{3}{2} \log \frac{m_t^2} {M_{H^\pm}^2}\non
\\
\Delta^+_{bt} = \frac{9}{4} + \frac{3}{2} \log \frac{m_b^2} {M_{H^\pm}^2}
\label{eq:d+}
\end{eqnarray}
The large logarithms can be mapped into the running quark masses in the usual
way. Adopting the $\overline{\rm MS}$ mass at the scale of the Higgs mass
the bulk of the next--to--leading order correction is automatically included
in this limit and the QCD corrections approach the common chirally invariant
factor $\Delta_{\rm QCD}$, eq.(\ref{eq:hqq}). \s

In the small mass limit, the non--logarithmic term of the factor
$\Delta^-_{tb}$
\begin{eqnarray}
\Delta^-_{tb} = 3+ \frac{3}{2} \log \frac{m_t^2} {M_{H^\pm}^2}+
\frac{3}{2} \log \frac{m_b^2} {M_{H^\pm}^2}
\label{eq:d-}
\end{eqnarray}
differs from the corresponding term in $\Delta^+_{ij}~(i,j=t,b)$,
eq.(\ref{eq:d+}). However, this is still in accordance with chiral symmetry
since the correction eq.(\ref{eq:d-}) is of subleading order in the small quark
mass expansion.

%\newpage
\vspace*{3cm}

\nn
{\large \bf Figure Captions}
%           ===============

\bigskip

\nn {\bf Fig.~1:}
(a) Generic Feynman diagrams for the decay processes $H \ra b\bar{b},
c \bar{c}$ and $H\to b\bar bg, c\bar c g$. (b) Feynman diagrams for
$H \ra gg$ decays and final--state gluon splitting into quarks.

\bigskip

\nn {\bf Fig.~2:}
Branching ratios of the \SM Higgs boson including the uncertainties from
the quark masses and the QCD coupling $\alpha_s$. The resulting
errors of the branching ratios are presented as shaded bands. The curves
labeled $b\overline{b}$ and $c\overline{c}$ are the inclusive decay
widths; they account for {\it all} final
states including $c$ and $b$ quarks. The curve labeled $gg$ corresponds
accordingly to gluon and light-quark final states only.

\bigskip

\nn {\bf Fig.~3:}
Branching ratios of the \MSSM~Higgs bosons $h$, $H$, $A$, $H^\pm$ [(a)..(d)],
including uncertainties from the quark masses $m_b, m_c, m_s$ and the strong
coupling $\alpha_s$. In the widths, the three--body channels \cite{susycoup}
have been included. The top quark mass is fixed at $M_t = 176~\GeV$ for the
shaded error bands and the error bars shown below the 200 \GeV~mark in figure
(b). The additional uncertainty due to the top mass is marked by error bars
in the figures (b) [at $M_H > 200~\GeV$], (c) and (d). In
figure (a) the curves for the upper and lower limit of the top mass band are
presented separately, using the average values of the other quark masses and
of the strong coupling $\alpha_s$ for the sake of clarity. The labels follow
the definitions in Fig.2; i.e. the branching ratios are classified according to
the inclusive hadronic final states with [labels $b\bar b, c\bar c$] and
without heavy quarks [label $gg$].


\vspace*{2cm}

\begin{thebibliography}{99}
\bibitem{Z1} P.M.~Zerwas, Proceedings of the International Conference on High
Energy Physics, Marseille 1993; A.~Djouadi, Int. Mod. Phys. {\bf A10} (1995) 1.
\bibitem{Z1a} H.M. Georgi, S.L. Glashow, M.E. Machacek and D.V Nanopoulos,
Phys. Rev. Lett. {\bf 40} (1978) 692.
\bibitem{R14A} M.~Spira, A.~Djouadi, D.~Graudenz and P.M.~Zerwas,
Nucl.~Phys.~{\bf B453} (1995) 17.
\bibitem{Z1b} A. Djouadi, J. Kalinowski and P. M. Zerwas, Z. Phys. {\bf C54}
(1992) 255.
\bibitem{Z1c} Z. Kunszt, Nucl. Phys. {\bf B247} (1984) 339;
J.F. Gunion, Phys. Lett. {\bf B253} (1991) 269.
\bibitem{Z1d} M.D.~Hildreth, T.L.~Barklow and D.L.~Burke, Phys.~Rev.~
             {\bf D49} (1994) 3441.
\bibitem{GKW} E.~Gross, B.A.~Kniehl and G.~Wolf, Z.~Phys.~{\bf C63} (1994) 417;
(E) {\bf C66} (1995) 321.
\bibitem{Z3} E. Braaten and J.P. Leveille, Phys. Rev. {\bf D22} (1980) 715;
A.L.~Kataev and V.T.~Kim, Mod.~Phys.~Lett.~{\bf A9} (1994) 1309;
L.R.~Surguladze, Phys. Lett. {\bf 341} (1994) 61.
\bibitem{Z3A} S.G.~Gorishny, A.L.~Kataev, S.A.~Larin and L.R.~Surguladze,
Mod.~Phys.~Lett.~{\bf A5} (1990) 2703; Phys.~Rev.~{\bf D43} (1991) 1633.
\bibitem{chetyr} K.G.~Chetyrkin, J.H.~K\"uhn and A.~Kwiatkowski, Proceedings
of the Workshop ''QCD at LEP'', Aachen 1994;
K.G.~Chetyrkin and A.~Kwiatkowski, Report LBL--37269.
%(May 1995).
\bibitem{helw} J.~Fleischer and F.~Jegerlehner, Phys.~Rev.~{\bf D23} (1981)
2001;
D.Yu.~Bardin, B.M.~Vilenski\u\i~and P.Kh.~Khristova,
Sov.~J.~Nucl.~Phys.~{\bf 53} (1991) 152;
A.~Dabelstein and W.~Hollik, Z.~Phys.~{\bf C53} (1992) 507;
B.A.~Kniehl, Nucl.~Phys.~{\bf B376} (1992) 3.
\bibitem{Z5} N.~Gray, D.J.~Broadhurst, W.~Grafe and K.~Schilcher,
Z.~Phys.~{\bf C48} (1990) 673.
\bibitem{Z4} S.~Narison, Phys.~Lett.~{\bf B341} (1994) 73.
\bibitem{Z6A} S. Bethke, Proceedings of the Workshop ``QCD 94", Montpellier
1994.
\bibitem{R11A} J. Ellis, M.K. Gaillard and D.V. Nanopoulos, Nucl. Phys.
{\bf B106} (1976) 292.
\bibitem{Hgg} T. Inami, T. Kubota and Y. Okada, Z. Phys. {\bf C18} (1983) 69;
A.~Djouadi, M.~Spira and P.M.~Zerwas, Phys. Lett. {\bf B264} (1991) 440.
\bibitem{BLM} S.J.~Brodsky, G.Peter~Lepage and P.B.~Mackenzie,
              Phys.~Rev.~{\bf D28} (1983) 228.
\bibitem{R8A} A. Menzione, Proceedings, International Conference on High
Energy Physics, Brussels 1995.
\bibitem{carena} M.~Carena, J.~Espinosa, M.~Quiros and C.~Wagner,
Phys.~Lett.~{\bf B355} (1995) 209.
\bibitem{susycoup} A. Djouadi, J. Kalinowski and P. M. Zerwas,
Report DESY 95--211.
\bibitem{Drehi} M. Drees and K. Hikasa, Phys. Lett. {\bf B240} (1990) 455.
\bibitem{R15} A. Djouadi and P. Gambino, Phys. Rev. {\bf D51} (1995) 218.

\end{thebibliography}
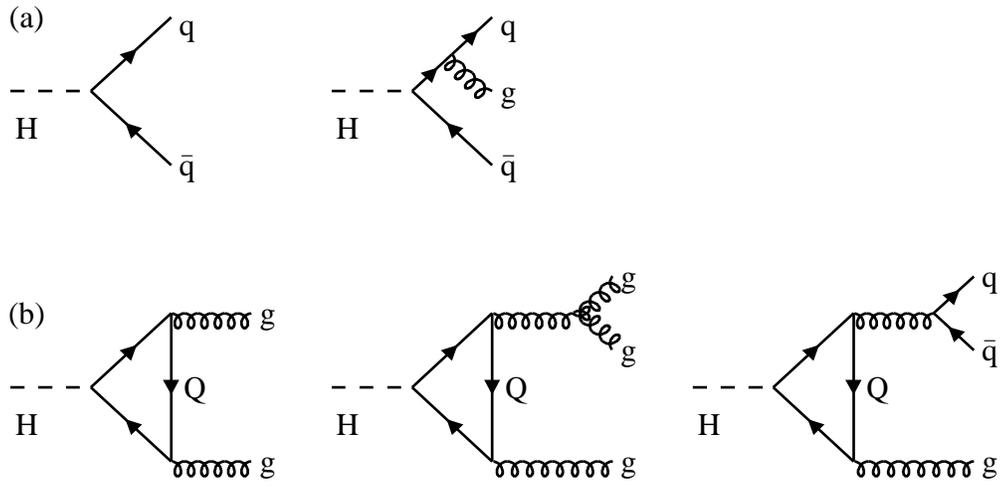
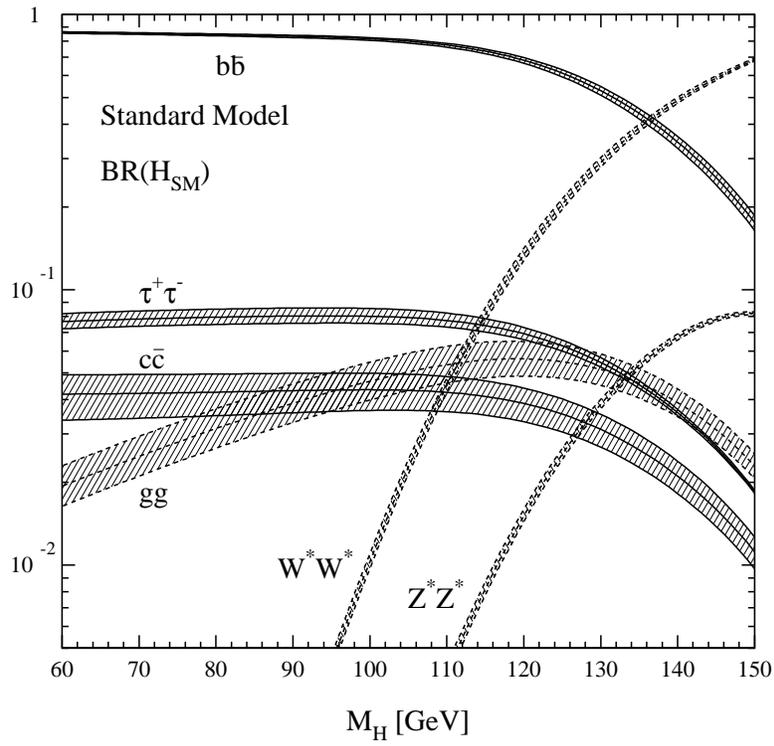
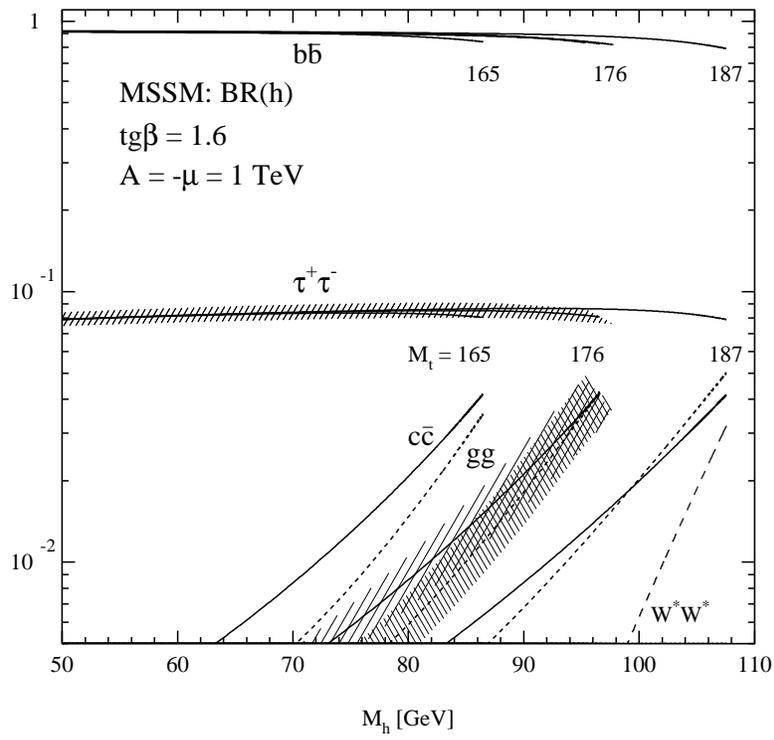
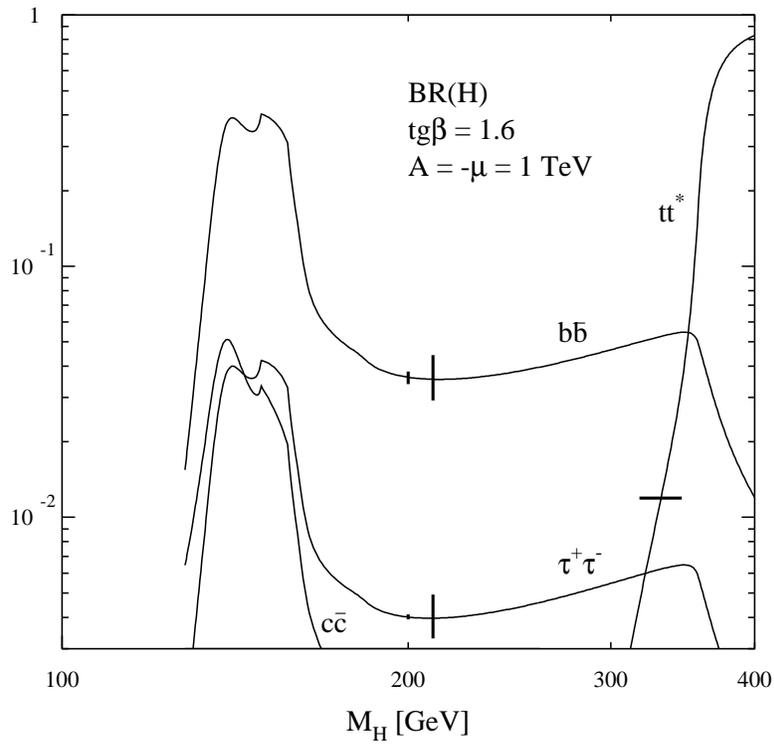
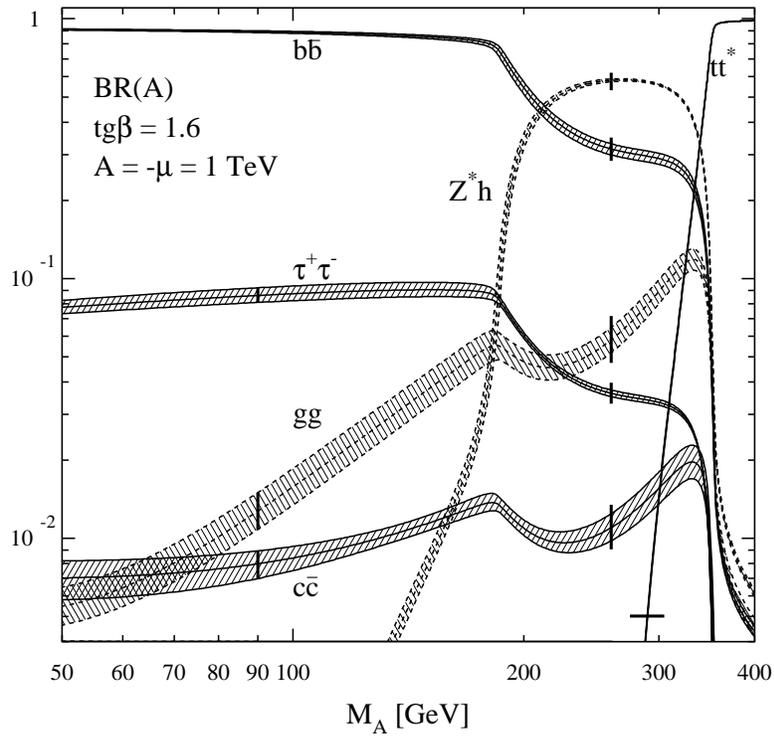
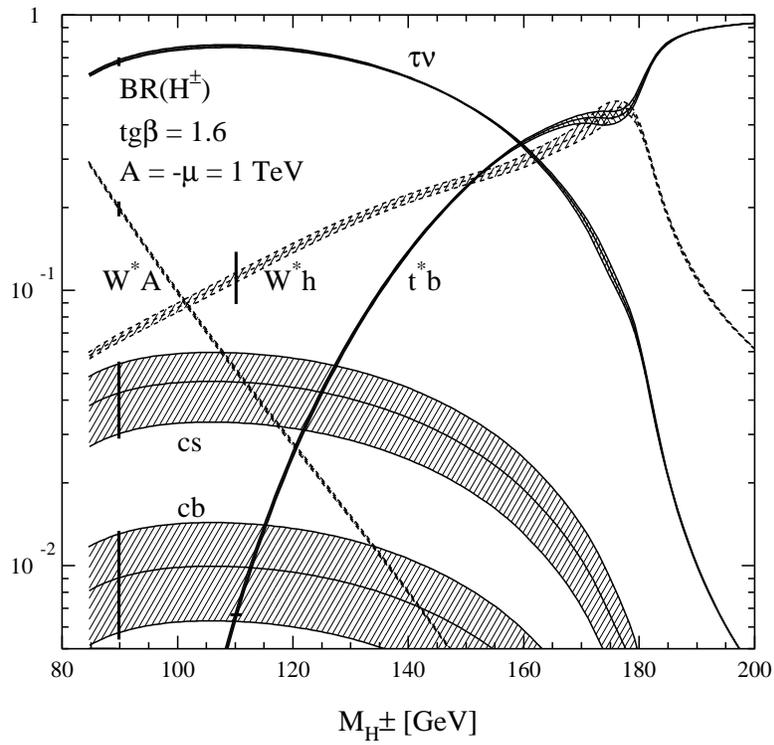
\end{document}